\documentclass{article}

\usepackage[square,numbers]{natbib}
\usepackage[final]{neurips_2022}

\usepackage[utf8]{inputenc} 
\usepackage[T1]{fontenc}    
\usepackage{hyperref}       
\usepackage{url}            
\usepackage{booktabs}       
\usepackage{amsfonts}       
\usepackage{nicefrac}       
\usepackage{microtype}      
\usepackage{xcolor}         

\usepackage{amsmath,amsfonts,bm}

\def\eqref#1{equation~\ref{#1}}

\def\1{\bm{1}}

\DeclareMathAlphabet{\mathsfit}{\encodingdefault}{\sfdefault}{m}{sl}
\SetMathAlphabet{\mathsfit}{bold}{\encodingdefault}{\sfdefault}{bx}{n}

\newcommand{\R}{\mathbb{R}}

\usepackage{graphicx} 
\usepackage{xspace}
\usepackage{multirow}
\usepackage{subcaption}

\usepackage{xr}
\makeatletter
\newcommand*{\addFileDependency}[1]{
  \typeout{(#1)}
  \@addtofilelist{#1}
  \IfFileExists{#1}{}{\typeout{No file #1.}}
}
\makeatother

\newcommand*{\myexternaldocument}[1]{%
    \externaldocument{#1}%
    \addFileDependency{#1.tex}%
    \addFileDependency{#1.aux}%
}

\myexternaldocument{main_supplementary}

\newcommand{\mc}[3]{\multicolumn{#1}{#2}{#3}}

\usepackage{macros}

\title{Spherical Channels for Modeling Atomic Interactions}

\author{
  \textbf{C. Lawrence Zitnick}$^1$, Abhishek Das$^1$, Adeesh Kolluru$^2$, Janice Lan$^1$, Muhammed Shuaibi$^2$, \\
  \textbf{Anuroop Sriram}$^1$, \textbf{Zachary Ulissi}$^2$, \bf{Brandon Wood}$^1$ \\
  $^1$ Fundamental AI Research at Meta AI \\
  $^2$ Carnegie Mellon University
}

\begin{document}

\maketitle

\begin{abstract}
Modeling the energy and forces of atomic systems is a fundamental problem in computational chemistry with the potential to help address many of the world's most pressing problems, including those related to energy scarcity and climate change. These calculations are traditionally performed using Density Functional Theory, which is computationally very expensive. Machine learning has the potential to dramatically improve the efficiency of these calculations from days or hours to seconds. 

We propose the Spherical Channel Network (SCN) to model atomic energies and forces. The SCN is a graph neural network where nodes represent atoms and edges their neighboring atoms. The atom embeddings are a set of spherical functions, called spherical channels, represented using spherical harmonics. We demonstrate, that by rotating the embeddings based on the 3D edge orientation, more information may be utilized while maintaining the rotational equivariance of the messages. While equivariance is a desirable property, we find that by relaxing this constraint in both message passing and aggregation, improved accuracy may be achieved. We demonstrate state-of-the-art results on the large-scale Open Catalyst 2020 dataset in both energy and force prediction for numerous tasks and metrics. 
\end{abstract}

\section{Introduction}
\label{sec:intro}
Modeling the properties of atomic systems is a foundational challenge in computational chemistry and critical to advancing technologies across numerous application domains. Notable applications include drug discovery~\cite{ramakrishnan2014quantum,senior2020improved} and the design of new catalysts for renewable energy storage to help in addressing climate change~\cite{zitnick2020introduction, rolnick2019tackling}. For catalyst discovery, new materials are currently evaluated using Density Functional Theory (DFT) that can estimate atomic energies and forces but is computationally very expensive; taking hours or days to evaluate a single material. Machine Learning (ML) has the potential to approximate DFT and dramatically speed up these calculations, allowing for high throughput screening of new materials to help address some of the world's most pressing challenges.

Our goal is to approximate DFT calculations using ML. An ML model takes as input a set of atom positions and their atomic numbers. As outputs, a model calculates the structure's energy (or other properties) and the per-atom forces, i.e., the forces exerted on each atom by the other atoms. A common approach to this problem is to use Graph Neural Networks (GNNs) \cite{gori2005new,zhou2020graph} where each node represents an atom and the set of nearby atoms as edges~\cite{schutt2017quantum,gilmer2017neural,jorgensen2018neural,schutt2017schnet,schutt2018schnet,xie2018crystal,qiao2020orbnet,klicpera2020directional}. 

A key challenge in network design is balancing the use of model constraints. When modeling atomic systems, a common constraint is SO(3) rotation equivariance  \cite{weiler20183d,batzner20223,anderson2019cormorant,thomas2018tensor,satorras2021n,schutt2021equivariant}, \ie, if the atomic system is rotated, the energies should remain constant and the atomic forces should similarly rotate. While this provides strong priors on the model to help in generalization, especially for smaller datasets\cite{chmiela2017machine,ramakrishnan2014quantum}, it can result in limiting the expressiveness of the network due to restrictions on non-linear transformations for equivariant models \cite{batzner20223,thomas2018tensor,satorras2021n,schutt2021equivariant}, or limiting interactions to pairs \cite{schutt2017schnet,schutt2018schnet,xie2018crystal}, triplets \cite{klicpera2020directional} or quadruplets of atoms \cite{liu2021spherical,klicpera2021gemnet,gasteiger2022graph} for invariant models. Alternatively, a non-equivariant model \cite{hu2021forcenet} can provide more freedom to the model, but lead to the model needing to learn approximate equivariance through methods such as data augmentation \cite{shorten2019survey}. To draw an analogy with image detection, the use of a CNN \cite{lecun1998gradient} provides translation equivariance and removes the need for the network to learn how to detect the same object at different locations. However, most CNNs are not equivariant to scale or rotation \cite{weiler2018learning,weiler20183d}, but are still effective in learning approximate equivariance through rotation and scale diversity in the training data. What is the analogous balance of constraints for modeling atomic systems?

In this paper, we propose a GNN \cite{gori2005new,zhou2020graph} that balances the use of model constraints to aid in generalization while providing the network with the flexibility to learn accurate representations. We introduce the Spherical Channel Network (SCN) that explicitly models relative orientations of all neighboring atoms; information that is critical to accurately predicting atomic properties. Each node's embedding is a set of functions defined on the surface of a sphere $(S^2 \rightarrow \R)$. The functions are represented using spherical harmonics, similar to approaches that build strictly equivariant models~\cite{thomas2018tensor,anderson2019cormorant,batzner20223,esteves2018learning}. During message passing, angular information between atoms is conveyed by rotating or steering \cite{brandstetter2021geometric} the embeddings based on each edge's orientation~\cite{shuaibi2021rotation,liu2021spherical}. We identify an expanded set of spherical harmonic coefficients that are invariant to rotation, which can provide rich information while maintaining a message's rotation equivariance. In addition, we demonstrate that if the equivariance constraint is relaxed, improved performance can be achieved by using additional coefficients. We further improve the expressivity of the network by performing a pointwise non-linear function, which is only approximately equivariant, on the embeddings during message aggregation.  

We demonstrate our Spherical Channel Network on the large-scale Open Catalyst 2020 (OC20) dataset \cite{OC20}, which contains atomic structures useful for numerous applications important to addressing climate change. State-of-the-art results are achieved for atomic force and initial structure to relaxed energy prediction with improvements of $8\%-11\%$. In addition, we demonstrate our model is more sample efficient as compared to other state-of-the-art models.

\section{Approach}
\label{sec:approach}

\begin{figure}
\begin{subfigure}{.5\textwidth}
  \centering
  \includegraphics[width=.9\linewidth]{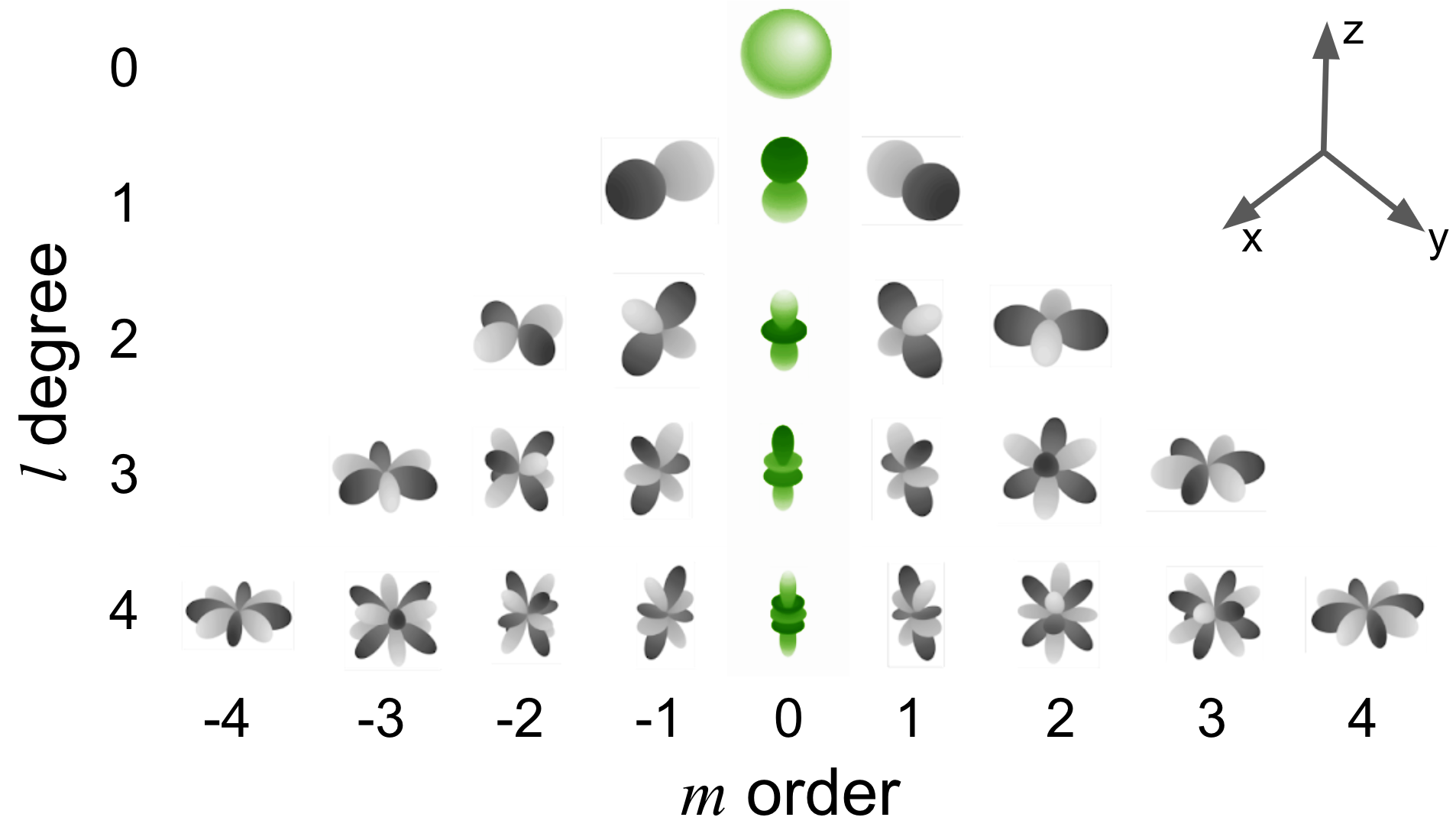}
  \caption{}
\end{subfigure}%
\begin{subfigure}{.5\textwidth}
  \centering
  \includegraphics[width=.98\linewidth]{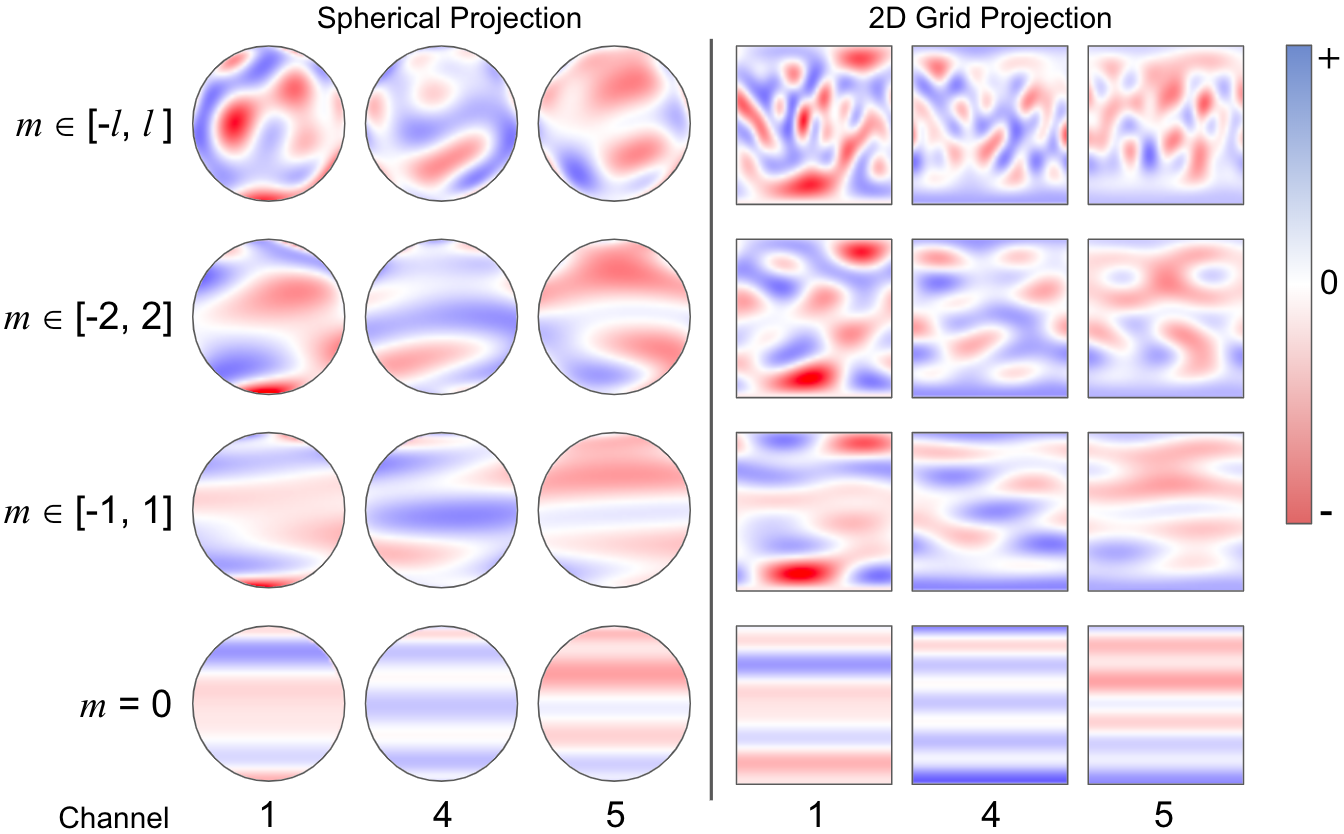}
  \caption{}
\end{subfigure}
\caption{(a) Illustration of spherical harmonics up to $l=4$. Note the $m=0$ bases are symmetric about the z-axis (center green). (b) Visualization of 3 spherical channels with $l=8$ when varying the number of orders: all, $m \in [-2,2]$, $m \in [-1,1]$, $m = 0$. Note how the resolution decreases with fewer $m$ until values are constant for a given z with $m=0$. Spherical projections are shown on the left (only half of the channel is visible) and a 2D grid projection (polar and azimuthal) of the same channels is shown on the right.}
\label{fig:sphharm}
\end{figure}

Given an atomic structure with $n$ atoms, our goal is to predict the structure's energy $E$ and the per-atom forces $\bm{f}_i$ for each atom $i \in n$. These values are estimated using a Graph Neural Network (GNN) \cite{gori2005new,zhou2020graph} where each node represents an atom and edges represent nearby atoms. As input, the network is given the distance $d_{ij}$ between atoms $i$ and $j$, and each atom's atomic number $a_i$. The neighbors $N_i$ for an atom $i$ are determined using a fixed distance threshold, or by picking a fixed number of closest atoms.

\begin{figure}
\begin{subfigure}{.55\textwidth}
    \centering
      \includegraphics[width=.98\linewidth]{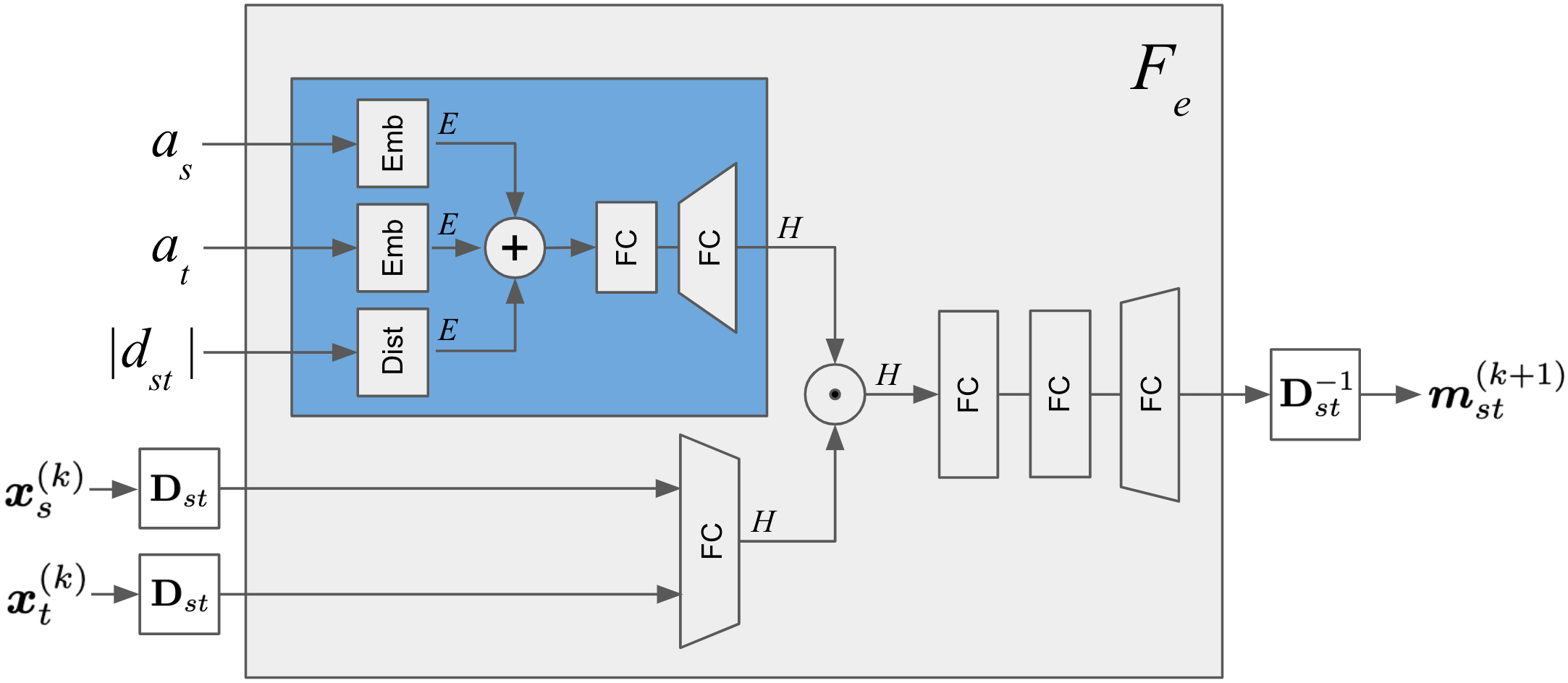}
      \caption{}
\end{subfigure}%
\begin{subfigure}{.45\textwidth}
    \centering
      \includegraphics[width=.9\linewidth]{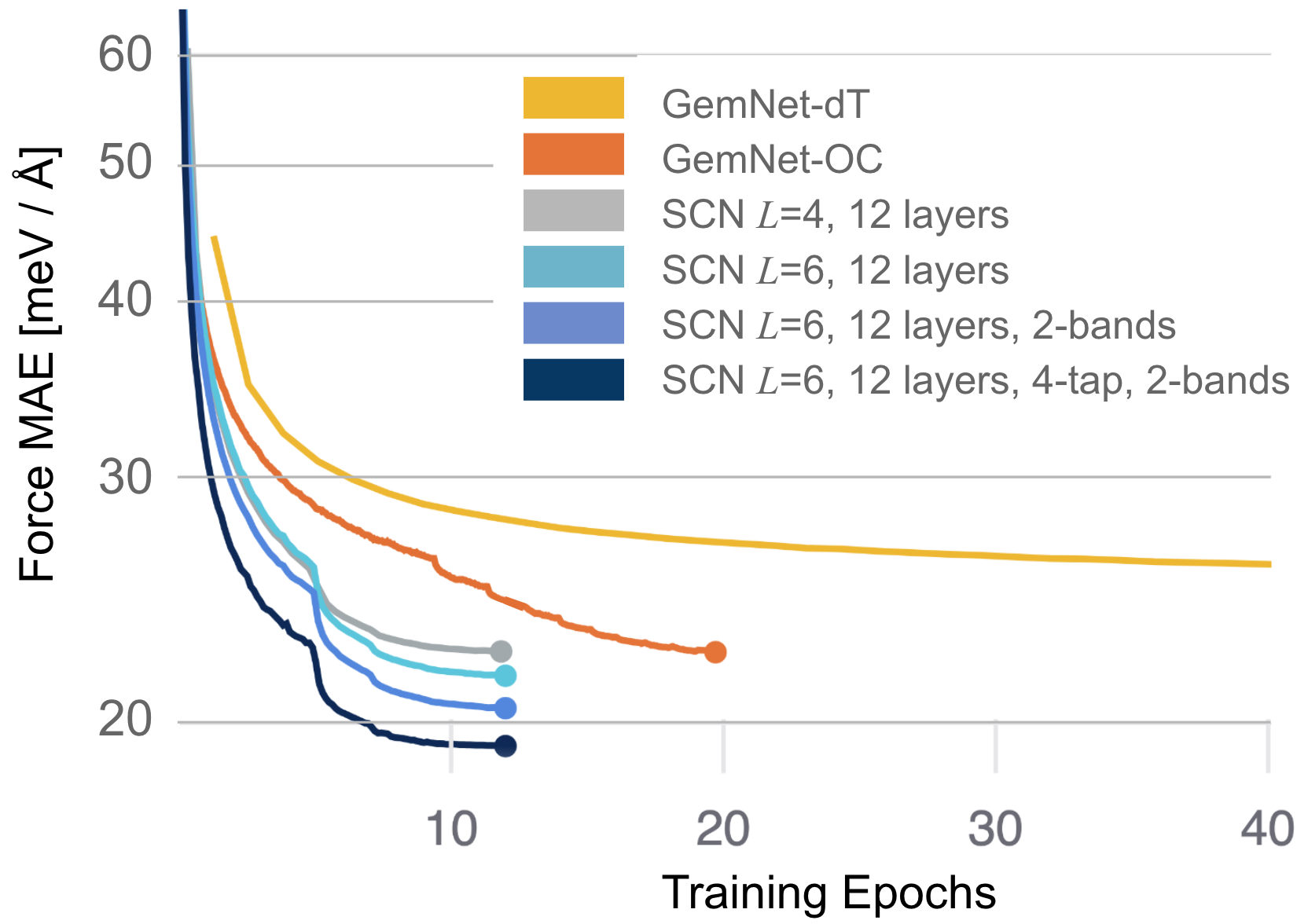}
      \caption{}
\end{subfigure}%
\caption{(a) Block diagram of message passing function $\bm{F}_e$ in Equation \ref{eqn:msg} for source atom $s$ to target atom $t$. The atomic numbers $a_s$ and $a_t$, distance between the atoms $|d_{st}|$, and embeddings $\bm{x}_s^{(k)}$ and $\bm{x}_t^{(k)}$ are given as input. (b) Training curves for SCN and GemNet models for force MAEs per epoch evaluated on a 30k subset of the validation ID dataset. Note how the SCN model is significantly more sample efficient during training.}
\label{fig:msg}
\end{figure}


\subsection{Node Embeddings}

The angular or relative orientation information between atoms is critical to accurately modeling atomic properties such as energies and forces \cite{klicpera2020directional, klicpera2021gemnet}. Inspired by this, our node embedding models the angular information from all neighboring atoms using spherical functions. Each node $i$'s embedding $s_{i}$ is a set of $C$ functions or channels represented on a sphere $(S^2 \rightarrow \R)$, whose argument is a 3D unit vector indicating the orientation. That is, $s_{ic}(\hat{\bm{r}})$ is the value of channel $c$ for node $i$ for some orientation $\hat{\bm{r}}\in \R^3$. Since the spherical channels contain orientation information over the entire sphere, the network may reason about geometric information for all neighboring atoms and not just atom pairs, triplets, etc. Each spherical channel $c$ may be represented using several different approaches, such as discrete 2D grids sampled over the sphere, or spherical harmonics. As we describe later, we use spherical harmonics due to their property of being SO(3)-equivariant (3D rotation equivariant). 

Using real spherical harmonics, a function $s_c(\hat{\bm{r}})$ defined on the sphere is represented by a set of weighted spherical harmonics $Y_{lm}$ with $s_c(\hat{\bm{r}}) = \sum_{l,m} \bm{x}_{lmc}Y_{lm}(\hat{\bm{r}})$ where $l$ and $m$ are the degree and order of the basis functions with $m \in [-l, l]$ and $l \in L$. We refer to the spherical harmonic coefficients, $\bm{x}_{lmc}$, as the coefficients of the spherical channel. For every degree $l$ there exists $2l + 1$ spherical harmonics (see Figure \ref{fig:sphharm}(a) for an illustration of functions for $l \le 4$), which results in a total of $(L+1)^2$ basis functions and coefficients up to degree $L$. Therefore for a maximum degree of $L$ with $C$ channels, $\bm{x}$ has size $(L+1)^2 \times C$.

The spherical channels are updated by the GNN through message passing for $K$ layers to obtain the final node embeddings $S^{(K)}$. $S^{(0)}$ is initialized from an embedding based on the atom's atomic number $a_i$ for $l=0$ coefficients and the $l \neq 0$ coefficients are set to zero.  The nodes' embeddings are updated by first calculating a set of messages $m_{ij}$ for each edge, which are then aggregated at each node. Finally, the energy and forces are estimated from $S^{(K)}$. We describe each of these steps in turn.

\subsection{Message Passing}

Given a target node $t$ and its neighbors $s \in N_t$ we want to update the embeddings $\bm{x}_t^{(k)}$ at iteration $k \in K$. The embeddings $\bm{x}^{(k)}_t$ are a set of spherical harmonic coefficients indexed by their degree $l$, order $m$, and channel $c$. An important and useful property of spherical harmonics is the function represented by the coefficients is steerable \cite{freeman1991design,brandstetter2021geometric}, \ie, it can be rotated in 3D space using a linear transformation of the coefficients. Specifically, for a 3D rotation matrix $\mathbf{R}$ there exists a matrix called a Wigner D-matrix $\mathbf{D}^l$ of size $(2l+1 \times 2l+1)$ that rotates the coefficients of degree $l$ by the rotation $\mathbf{R}$. If $\bm{x}_{ilc}$ are all coefficients across orders $m$ for node $i$ of degree $l$ and channel $c$, then for any 3D rotation $\mathbf{R}$ there exists a $\mathbf{D}^l$ that for all orientations $\hat{\bm{r}}$: 
\begin{equation}
\bm{x}_{ilc}\cdot Y_l(\mathbf{R}\hat{\bm{r}}) =  (\mathbf{D}^l\bm{x}_{ilc}) \cdot Y_l(\hat{\bm{r}}).
\end{equation}
When calculating the message $m_{st}$ from atom $s$ to atom $t$, we want to use the information contained in both $\bm{x}_s^{(k)}$ and $\bm{x}_t^{(k)}$ given the context of the edge's orientation $\bm{\hat{d}}_{st}$, $\bm{\hat{d}}_{st} = \bm{d}_{st} / |\bm{d}_{st}|$. We do this by rotating the embeddings by $\mathbf{R}_{st}$ for which $\mathbf{R}_{st}\bm{\hat{d}}_{st} = [0,0,1]^{\top}$, \ie, the direction of the edge's unit vector $\bm{\hat{d}}_{st}$ is aligned with the z-axis. Thus the orientation of the atoms with respect to each other is implicit in the rotated embeddings. This simplifies the task of the network since it only needs to learn the relationship between atoms that are aligned along the z-axis and not an arbitrary rotation. After calculating the messages using a neural network $\bm{F}_e$, they are rotated back to the global coordinate frame:
\begin{equation}
\bm{m}_{st}^{(k+1)}(\bm{x}^{(k)}_s, \bm{x}^{(k)}_t, \bm{d}_{st}, a_s, a_t) = \mathbf{D}_{st}^{-1} \bm{F}_e (\mathbf{D}_{st} \bm{x}^{(k)}_s, \mathbf{D}_{st} \bm{x}^{(k)}_t, |\bm{d}_{st}|, a_s, a_t),
\label{eqn:msg}
\end{equation}
where the matrices $\mathbf{D}_{st}$ and $\mathbf{D}_{st}^{-1}$ perform the rotation on the embeddings' coefficients corresponding to the 3D rotation matrices $\mathbf{R}_{st}$ and $\mathbf{R}_{st}^{-1}$ respectively. To simplify notation, we assume the coefficients in $\bm{x}_s$ across degree $l$ and order $m$ are flattened, and $\mathbf{D}_{st} \in \R^{(L+1)^2} \times \R^{(L+1)^2}$ is a block diagonal matrix constructed from the set of Wigner D-matrices across degrees, $l \in L$. $\mathbf{D}_{st}^{-1}$ is the inverse of $\mathbf{D}_{st}$, with $\mathbf{D}_{st}^{-1} = \mathbf{D}_{st}^{\top}$. In addition to the rotated embeddings, $\bm{F}_e$ is also provided input edge information that is invariant to rotations; $|\bm{d}_{st}|$ is the magnitude of the distance between atoms $s$ and $t$, and $a_s$ and $a_t$ their respective atomic numbers. Note that the rotation matrix $\mathbf{R}_{st}$ is not unique, since the roll rotation around the vector $\bm{d}_{st}$ is not specified and is randomly sampled during training. The implications of this are discussed below in Section \ref{sec:equiv}. 

The message function $\bm{F}_e$ is computed using a neural network as illustrated in Figure \ref{fig:msg}(a). The atomic numbers $a_s$ and $a_t$ are used to look up two independent embeddings of size $E=128$, and a set of 1D basis functions are used to represent $|\bm{d}_{st}|$ using equally spaced Gaussians every 0.02 \AA~from 0 to the 8 \AA~with $\sigma = 0.04$ followed by a linear layer with $E$ outputs. Their values are added together and passed through a neural network to produce a vector of size $H$. The rotated embeddings $\mathbf{D}_{st} \bm{x}^{(k)}_s$ and $\mathbf{D}_{st} \bm{x}^{(k)}_t$ are concatenated and passed through a single layer neural network to produce another vector of size $H$. These two are multiplied together to combine the edge information with the information contained in the rotated embeddings. Two more fully connected layers are performed with each followed by a SiLU non-linear activation function \cite{hendrycks2016gaussian}. Finally, a single linear layer is used to expand the output to the original size of the embeddings.

\subsubsection{Message Equivariance}
\label{sec:equiv}

For the network to be equivariant to rotations, our message function (Equation (\ref{eqn:msg})) needs to be equivariant. In general, this is not the case since the matrix $\mathbf{D}_{st}$ is not unique, \ie, the roll rotation around the vector $\bm{d}_{st}$ is not specified, and in practice it is randomly chosen. The roll rotation corresponds to a rotation about the z-axis after rotating the coefficients $\bm{x}_s^{(k)}$ and $\bm{x}_t^{(k)}$ by $\mathbf{D}_{st}$. Due to this, all of the rotated coefficients will vary based on the random roll rotation chosen, except the $m=0$ coefficients that are symmetric about the z-axis; see Appendix \ref{sec:appx-sphharm} and the $m=0$ bases highlighted in green in Figure \ref{fig:sphharm}(a). If rotation equivariance is desired for messages, the input and output coefficients to $\bm{F}_e$ can be restricted to only those for which $m=0$. Since the other inputs $|\bm{d}_{st}|$, $a_s$ and $a_t$ to $\bm{F}_e$ are also invariant to rotations, $\bm{F}_e$ is invariant to rotations if only $m=0$ coefficients are used. The resulting messages $\bm{m}_{st}$ are equivariant to rotations once they are rotated back to the global coordinate frame using  $\mathbf{D}_{st}^{-1}$. While equivariance is a desirable property, as we demonstrate later, using $m \in [-1, 1]$ coefficients can help improve performance even though equivariance is not strictly enforced.

If we choose to use coefficients beyond just $m=0$ for $\bm{F}_e$ for message passing in Equation (2), two approaches may be taken. The first is to simply add the coefficients to the inputs and outputs of $\bm{F}_e$ and assume the neural network will learn a roughly equivariant mapping through the random sampling of the roll rotation. As we demonstrate in Appendix \ref{sec:appx-equiv}, the learned functions are indeed roughly equivariant. The second strategy takes a more direct approach to encouraging the network to learn equivariant mappings by taking advantage of the proprieties of the coefficients as they're rotated by $\phi$ about the z-axis. If $\mathbf{D}_{\phi}$ rotates a set $x$ of coefficients about the z-axis by $\phi$, the $m=0$ coefficients are constant as a function of $\phi$, while the $m \in \{-1, 1\}$ coefficients are sine and cosine functions of $\phi$ (see Appendix \ref{sec:appx-sphharm}):
\begin{equation}
    (\mathbf{D}_{\phi}\bm{x})_{(0)} = \bm{\gamma},
    \label{eqn:in0}
\end{equation}
\begin{equation}
    (\mathbf{D}_{\phi}\bm{x})_{(-1)} = \bm{\alpha} \sin(\phi + \bm{\beta}) = (\mathbf{D_{(\phi-\frac{\pi}{2})}}\bm{x})_{(1)},
    \label{eqn:in1}
\end{equation}
\begin{equation}
    (\mathbf{D}_{\phi}\bm{x})_{(1)} = \bm{\alpha} \cos(\phi + \bm{\beta}) = (\mathbf{D_{(\phi+\frac{\pi}{2})}}\bm{x})_{(-1)},
    \label{eqn:in2}
\end{equation}
for some set of vectors $\bm{\alpha}, \bm{\beta}$ and $\bm{\gamma}$ of size $L$. $\bm{x}_{(0)}$ are the $m=0$ coefficients and similarly for $\bm{x}_{(-1)}$ and $\bm{x}_{(1)}$ for the $m=-1$ and $m=1$ coefficients respectively. Similar proprieties hold for $m > 1$ and $m < -1$. We take advantage of these properties to encourage the output of the message block towards equivariance by computing $\bm{F}_{e}$ at multiple rotations $\phi$ about the z-axis:
\begin{equation}
\bm{F}_{e}^{\phi} = \bm{F}_e (\mathbf{D}_{\phi}\mathbf{D}_{st} \bm{x}^{(k)}_s, \mathbf{D}_{\phi}\mathbf{D}_{st} \bm{x}^{(k)}_t, |\bm{d}_{st}|, a_s, a_t),
\label{eqn:msgphi}
\end{equation}
For $m \in [-1, 1]$, we compute four samples or "taps" at $\phi \in \{0, \frac{1}{2}\pi, \pi, \frac{3}{2}\pi\}$, and combine them based on $m$ using: 

\begin{equation}
\begin{aligned}
\bm{m}_{st}^{(k+1)} & = & \mathbf{D}_{st}^{-1}\frac{1}{4} \left(\bm{F}_{e(0)}^{0} + \bm{F}_{e(0)}^{\frac{1}{2}\pi} + \bm{F}_{e(0)}^{\pi} + \bm{F}_{e(0)}^{\frac{3}{2}\pi}\right) +\ \\
& & \mathbf{D}_{st}^{-1}\frac{1}{4} \left(\bm{F}_{e(-1)}^{0} - \bm{F}_{e(1)}^{\frac{1}{2}\pi} - \bm{F}_{e(-1)}^{\pi} + \bm{F}_{e(1)}^{\frac{3}{2}\pi}\right) + \\
& & \mathbf{D}_{st}^{-1}\frac{1}{4} \left(\bm{F}_{e(1)}^{0} + \bm{F}_{e(-1)}^{\frac{1}{2}\pi} - \bm{F}_{e(1)}^{\pi} - \bm{F}_{e(-1)}^{\frac{3}{2}\pi}\right)
\end{aligned}
\label{eqn:taps}
\end{equation}
where $\bm{F}_{e(0)}^{\phi}$, $\bm{F}_{e(-1)}^{\phi}$ and $\bm{F}_{e(1)}^{\phi}$ are the output coefficients for $m=0$,~$m=-1$ and $m=1$ respectively. The top line of Equation \ref{eqn:taps} is simply taking the average, since the $m=0$ coefficients should be constant regardless of $\phi$. Similarly for $m \in \{-1, 1\}$, the second and third lines average 4 values that should be equal if Equations \ref{eqn:in1} and \ref{eqn:in2} hold, e.g., $\bm{\alpha}\sin(\phi + \bm{\beta}) = -\bm{\alpha}\cos(\phi + \bm{\beta} + \pi/2)$. If higher $m$ are desired, similar calculations can be performed. However, a larger number of taps will be needed, \eg, $m\in[-2,2]$ requires 8 taps separated by $\frac{1}{4}\pi$ radians. 

The number of coefficients used during message passing when calculating $\bm{F}_e$ may be varied based on the edge properties to reduce the memory required by the network. For instance, atoms that are far away from each other may not need the same resolution of the spherical functions as those close together, so a lower maximum degree $L$ may be used. By training several message passing networks with varying degrees for different ranges of edge distances, memory usage can be reduced while not impacting accuracy. Similarly, the range of $m$ can typically be truncated at $[-1,1]$ or $[-2,2]$ since their observed utility to the network is substantially reduced for values of $m$ greater than 2 (or less than -2). See Figure \ref{fig:sphharm}(b) for examples of the spherical channels when the range of $m$ is reduced. Note that while the number of coefficients used by $\bm{F}_e$ may be reduced, the node embeddings maintain their original size based on the maximum $L$, so they may be able to aggregate information from all neighboring atoms at different distances and angles.

\subsection{Message Aggregation}

At this point in the network, the per edge messages are computed and rotated back to the global coordinate frame. Ideally, to allow for complex interactions between the messages, a non-linear function would be applied after summing all messages directed at a target node. If we did this by naively passing the summed coefficients $\bm{m}_{t}^{(k+1)} = \sum_s \bm{m}_{st}^{(k+1)}$ through a fully connected neural network, the network may have difficulties learning representations that are approximately rotation equivariant (a fundamental property of atom forces). Another option is to place constraints on the non-linearities performed to enforce equivariance \cite{kondor2018clebsch, brandstetter2021geometric, schutt2021equivariant, thomas2018tensor, batzner20223, batatia2022design}. 

\begin{figure}
  \centering
  \includegraphics[width=0.94\linewidth]{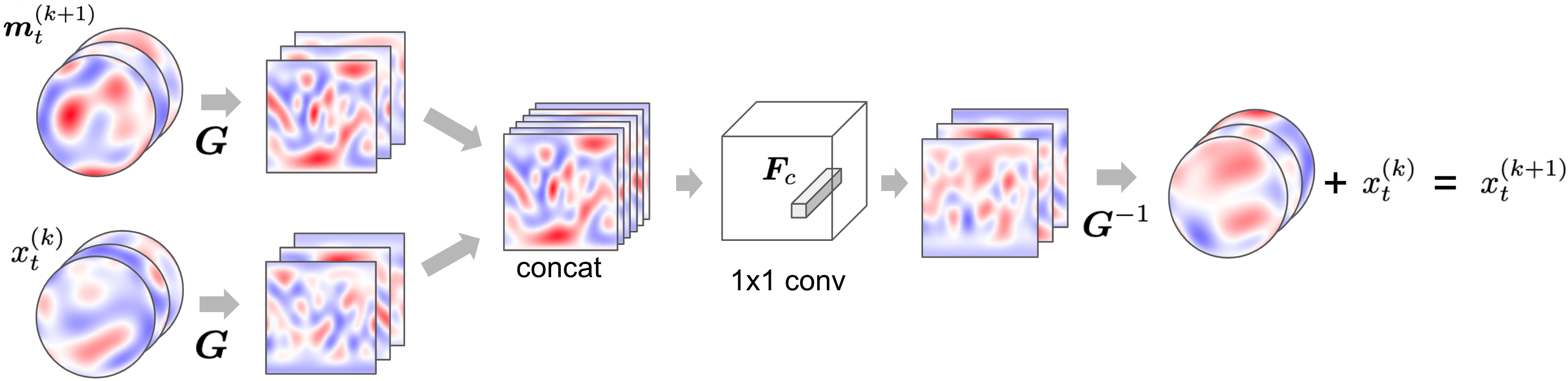}
  \caption{Illustration of message aggregation. The summed messages $\bm{m}_{t}^{(k+1)} = \sum_s \bm{m}_{st}^{(k+1)}$ and previous embedding $\bm{x}^{(k)}_t$ are converted from a spherical harmonic representation to a spherical grid representation using $\bm{G}$. The channels are concatenated and passed through a 3 layer $1\times1$ CNN. Each layer is followed by a SiLU activation. Finally, the channels are converted back to a spherical harmonic representation using $\bm{G}^{-1}$ and added to $\bm{x}^{(k)}_t$ to create $\bm{x}^{(k+1)}_t$.}
  \label{fig:grid}
  \end{figure}

We propose applying an unconstrained non-linear pointwise function on the sphere, i.e., a function that is applied at every orientation without knowledge of the orientation from which the point was sampled \cite{cohen2018spherical}. For our function, we use a non-linear neural network $\bm{F}_c$ ($\R^C \rightarrow \R^C$) that combines information across channels. In practice, $\bm{F}_c$ is applied at a discrete number of orientations, which results in a transformation that we numerically demonstrate (see Appendix \ref{sec:appx-equiv}) is approximately equivariant to rotations:
\begin{equation}
{\bm{x}}_t^{(k+1)} = \bm{x}_t^{(k)} + \bm{G}^{-1}\left(\bm{F}_c\left(\bm{G}(\bm{m}_{t}^{(k+1)}), \bm{G}(\bm{x}_t^{(k)})\right)\right),
\label{eqn:aggr}
\end{equation}
where $\bm{G}$ is a function that converts a spherical function represented by spherical harmonic coefficients, to one represented by point samples on a sphere. For $\bm{G}$ we use a spherical coordinate system (polar and azimuthal) to generate a 2D discrete representation of the spherical functions, see Figure \ref{fig:grid}. A $1 \times 1$ convolutional neural network $\bm{F}_c$ is applied to each discrete sample containing $C$ channels, and the result is then converted back to the spherical harmonic representation using $\bm{G}^{-1}$. This is analogous to having a 2D image represented in the frequency domain, converting it to the spatial domain using an inverse discrete Fourier transform, applying a transformation and converting back to the frequency domain. Since the same operation is applied to all orientations on the sphere, a transformation that is approximately equivariant (up to discrete sampling limitations) to rotations may be learned, see Appendix \ref{sec:appx-equiv} for a more detailed discussion.

\begin{figure}
  \centering
  \includegraphics[width=.95\linewidth]{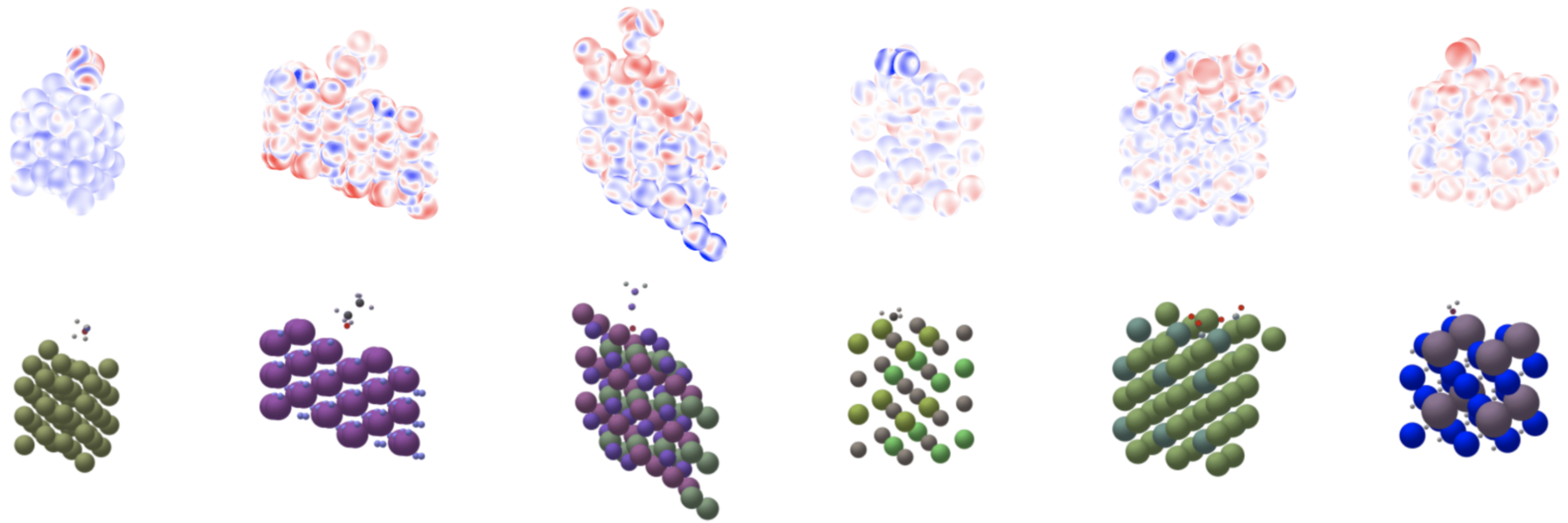}
  \caption{Illustration of different spherical channels for six structures (top). Blue indicates a positive value and red a negative value on the sphere. The spherical channels were sampled from the last layer before the output blocks using a model with 12 layers and $L=6$. Note, how some channels have a higher activation for adsorbates (darker colors), certain elements or the bottom or top of the surface. Illustrations of the structures are shown on the bottom with different colors indicating different elements.}
\label{fig:layer}
\end{figure}

A diagram of the network is shown in Figure \ref{fig:grid}. To provide additional information to the network, $\bm{x}_t^{(k)}$ is also converted to a spherical grid representation, concatenated with $\bm{G}(\bm{m}_{t}^{(k+1)})$ and provided as input to the neural network to compute the final updates to the coefficients $\bm{x}_t^{(k+1)}$ at iteration $k+1$. The network can learn circular functions on the sphere if varying resolutions of $\bm{x}^{(k)}$ are provided to the network, similar to Difference of Gaussian filters for 2D images \cite{lowe2004distinctive}. The resolution of $\bm{x}^{(k)}$ can be reduced by lowering the degree $L$ before transforming by $\bm{G}$. In our experiments, if multiple resolutions or bands are used, we use resolutions of degree $L$ and $L-1$. Three layers of $1 \times 1$ convolutions on $2C$ channels (or $4C$ if two bands are used) are performed with each followed by a SiLU non-linear activation function. To avoid aliasing, the spherical grid is sampled at a resolution of $2*(L + 1)$. See Figure \ref{fig:layer} for several example illustrations of spherical channels for different atomic structures.

\subsection{Energy and Forces Estimation}

We compute the energy $E$ by first estimating a per-atom energy using a pointwise function on the sphere and taking its integral over all possible orientations. The per-atom energy estimates are summed to obtain the overall system's energy:
\begin{equation}
E = \sum_i \int \bm{F}_{energy}\left(s_{i}^{(K)}(\hat{\bm{r}})\right)d\hat{\bm{r}},
\label{eqn:energy}
\end{equation}
where $\bm{F}_{energy}$ is a three layer fully connected neural network ($\R^C \rightarrow \R^1$) with SiLU activation functions. 

Forces may be calculated using two approaches: First, by calculating the gradients of the energy with respect to the atom positions. This approach enforces energy conservation, but due to the need to back-propagate the gradients is computationally much more expensive. The second approach computes the forces in a manner similar to energy prediction, which is computationally more efficient, but does not enforce energy conservation. In this approach, the forces are calculated by estimating a force magnitude $|\bm{f}| = \bm{F}_{force}(s_{i}^{(K)}(\hat{\bm{r}}))$ in every direction $\hat{\bm{r}}$ over the sphere. Integration is performed after multiplying the magnitude by the orientation $\hat{\bm{r}}$ to obtain directional vectors: 
\begin{equation}
\bm{f}_i = \int \hat{\bm{r}}\bm{F}_{force}\left(s_{i}^{(K)}(\hat{\bm{r}})\right)d\hat{\bm{r}},
\label{eqn:force}
\end{equation}
where $\bm{F}_{force}$ is a three layer fully connected neural network ($\R^C \rightarrow \R^1$). In practice, a discrete approximation of the integral is performed for equations (\ref{eqn:energy}) and (\ref{eqn:force}) using a set of 128 evenly distributed points on the sphere (see Appendix \ref{sec:sphere-sample}). Note that the application of a pointwise function of an arbitrary neural network at a finite number of discrete orientations is not invariant (energies) or equivariant (forces) to rotations. However in practice, numerically this operation does closely approximate these properties (see Appendix \ref{sec:equiv_efblocks}).

\section{Experiments}
\label{sec:experiments}
\begin{table*}[t]
    \centering
    \renewcommand{\arraystretch}{1.0}
    \setlength{\tabcolsep}{5pt}
    \resizebox{0.99\linewidth}{!}{
    \begin{tabular}{lccccc|cccc|cc}
       \mc{12}{c}{\ocd{} 2M Validation} \\
      \toprule
         & & & & &   & \mc{4}{c|}{\textbf{S2EF}} & \mc{2}{c}{\textbf{ IS2RE}} \\
        & &  & & & Samples / & Energy MAE  & Force MAE  & Force Cos  & EFwT & Energy MAE & EwT \\
        \textbf{Model} & &  & & & GPU sec. &  [meV] $\downarrow$ & [meV/\AA] $\downarrow$ &  $\uparrow$ & [\%] $\uparrow$ &  [meV] $\downarrow$ & [\%] $\uparrow$ \\

      \midrule
        Median & & &  && & &   &  & \\
      \midrule 
        SchNet \cite{schutt2018schnet} & & & & & & 1400 & 78.3 & 0.109 & 0.00 & - & - \\
        DimeNet++ \cite{klicpera_dimenetpp_2020} & & &  & & & 805 & 65.7 & 0.217 & 0.01 & - & - \\
        SpinConv \cite{shuaibi2021rotation} & & & & & & 406 & 36.2 & 0.479 & 0.13 & - & - \\
        GemNet-dT \cite{klicpera2021gemnet} & & & & & 25.8 & 358 & 29.5 & 0.557 & 0.61 & 438 & - \\
        GemNet-OC \cite{gasteiger2022graph} & & & & & 18.3 & 286 & 25.7 & 0.598 & 1.06 & 407 & - \\
        
        \midrule 
        & \textbf{$L$} & \textbf{\# layers} & \textbf{$H$} & \textbf{\# batch} & & & & & \\
        \midrule 
        {\model{} No rotation}  & 4 & 12 & 512 & 64 & 8.5 & 410 & 67.7 & 0.192 & 0.01 & - & - \\
        {\model{} No 1x1 conv}  & 6 & 12 & 1024 & 128 & 9.1 & 313 & 26.2 & 0.579 & 0.74 & - & - \\
        {\model{} $m=0$}  & 6 & 12 & 1024 & 128 & 10.4 & 307 & 26.5 & 0.588 & 0.83 & - & - \\
        
        {\model{} $m\in [-1,1]$}  & 6 & 12 & 1024 & 128 & 8.3 & 302 & 24.6 & 0.601 & 0.93 & - & - \\
        {\model{} $m\in [-2,2]$}  & 6 & 12 & 1024 & 96 & 6.9 & 301 & 23.4 & 0.612 & 1.01 & - & - \\
        {\model{} $m\in [-l,l]$}  & 4 & 12 & 512 & 64 & 7.6 & 297 & 24.6 & 0.595 & 0.92 & - & - \\
        \midrule
        {\model{} grad-forces}  & 4 & 12 & 512 & 128 & 1.2 & 307 & 26.2 & 0.573 & 0.81 & - & - \\
        {\model{} direct-forces}  & 4 & 12 & 512 & 128 & 12.1 & 303  & 25.3 & 0.592 & 0.87 & - & - \\
        \midrule
        {\model{}}  & 2 & 12 & 256 & 128 & 12.9  & 312 & 27.6 & 0.568 & 0.70 & - & - \\
        {\model{}}  & 4 & 12 & 512 & 128 & 12.1 & 303  & 25.3 & 0.592 & 0.87 & - & - \\
        
        {\model{}}  & 6 & 12 & 1024 & 128 & 8.3 & 302 & 24.6 & 0.601 & 0.93 & - & - \\
        {\model{}}  & 8 & 12 & 1024 & 64 & 5.9 & 300 & 23.2 & 0.620 & 1.15 & - & - \\
        \midrule 
        {\model{} }  & 6 & 12 & 1024 & 64 & 7.7 & 299 & 24.3 & 0.605 & 0.98 & - & - \\
        {\model{} 2-band}  & 6 & 12 & 1024 & 64 & 5.1 & 292 & 23.1 & 0.622 & 1.18 & - & - \\
        {\model{} 4-tap}  & 6 & 12 & 1024 & 64  & 3.7 & 296 & 22.7 & 0.638 & 1.27 & - & - \\
        {\model{} 4-tap 2-band}  & 6 & 12 & 1024 & 64 & 3.5 & \bf{279} & \bf{22.2} & 0.643 & \bf{1.41} & 371 & 11.0 \\
        \midrule
        {\model{} $m=0$}  & 6 & 16 & 1024 & 96 & 7.4 & 300 & 25.7 & 0.600 & 0.96 & 394 & 9.6 \\
        {\model{} $m=0$}  & 8 & 16 & 1024 & 64 & 4.8 & 296 & 25.3 & 0.608 & 1.01 & 389 & 9.9 \\
        {\model{}}  & 6 & 16 & 1024 & 96 & 5.9 &  287 & 22.8 & 0.623 & 1.22 & 371 & 10.5 \\
        {\model{}}  & 8 & 16 & 1024 & 96 & 3.5 & \bf{283}  & 22.7 & 0.627 & 1.22 & \bf{364} & \bf{11.3} \\ 
        {\model{} 4-tap}  & 6 & 16 & 1024 & 64 & 2.6 &  \bf{282} & \bf{22.2} & \bf{0.648} & 1.37 & 378 & 10.7 \\
        {\model{} 4-tap 2-band}  & 6 & 16 & 1024 & 64 & 2.3 & \bf{279}  & \bf{21.9} & \bf{0.650} & \bf{1.46} & 373 & \bf{11.0} \\
        
      \bottomrule
    \end{tabular}}
    \caption{Results on the \ocd~2M training dataset and ablation studies for \model{}~model variations. The validation results are averaged across the four OC20 Validation set splits. All SCN models are trained on 16 GPUs for 12 epochs with the learning rate reduced by 0.3 at 5, 7, 9, and 11 epochs, except SCN with $L=8$ and 16 layers that used 32 GPUs to obtain a larger batch size. Batch sizes vary based on the number of instances that can be fit in 32GB RAM.}
    \label{tab:comp-ablation}
    \vspace{-0.35cm}
\end{table*}

\begin{table*}[t]
    \centering
    \renewcommand{\arraystretch}{1.0}
    \setlength{\tabcolsep}{5pt}
    \resizebox{0.97\linewidth}{!}{
    \begin{tabular}{lrr|cccc|cc|c}
       \mc{10}{c}{\ocd{} Test} \\
      \toprule
         & & & \mc{4}{|c|}{\textbf{S2EF}} & \mc{2}{|c|}{\textbf{IS2RS}} & \textbf{IS2RE} \\
         & & \textbf{Train} & Energy MAE & Force MAE & Force Cos & EFwT & AFbT & ADwT & Energy MAE \\
        Model & \#Params & \textbf{time} & meV $\downarrow$ & [meV/\AA] $\downarrow$ & $\uparrow$ & [\%] $\uparrow$ & [\%] $\uparrow$ & [\%] $\uparrow$ & meV $\downarrow$ \\
      \midrule
        Median & -- & & 2258 & 84.4 & 0.016 & 0.01 & - & - & - \\
      \midrule
        & & & \mc{7}{c}{\textbf{Train \ocd~All}} \\
        SchNet \cite{schutt2018schnet,OC20}  & 9.1M & 194d &  540 & 54.7 & 0.302 & 0.00 & - & 14.4 & 764  \\
        PaiNN \cite{schutt2021equivariant}
            & 20.1M & 67d & 341 & 33.1 & 0.491 & 0.46 & 11.7 & 48.5 & 471 \\
        DimeNet++-L-F+E \cite{klicpera_dimenetpp_2020, OC20} & 10.7M & 1600d & 480 & 31.3 & 0.544 & 0.00 & 21.7 & 51.7 & 559 \\
        SpinConv (direct-forces) \cite{shuaibi2021rotation} & 8.5M & 275d & 336 & 29.7 & 0.539 & 0.45 & 16.7 & 53.6 & 437  \\
        GemNet-dT \cite{klicpera2021gemnet}
            & 32M & 492d & 292 & 24.2 & 0.616 & 1.20 & 27.6 & 58.7 & 400 \\
        GemNet-OC \cite{gasteiger2022graph}
            & 39M & 336d & \bf{233} & 20.7 & 0.666 & 2.50 & 35.3 & 60.3 & 355 \\
        \midrule
        {\model{} $L$=8 $K$=20}  & 271M & 645d & 244 & \bf{17.7} & \bf{0.687} & \bf{2.59} & \bf{40.3} & \bf{67.1} & \bf{330} \\
        \midrule
        & & & \mc{7}{c}{\textbf{Train \ocd~All + MD}} \\
        GemNet-OC-L-E \cite{gasteiger2022graph}
            & 56M & 640d & \bf{230} & 21.0 & 0.665 & 2.80 & - & - & - \\
        GemNet-OC-L-F \cite{gasteiger2022graph}
            & 216M & 765d & 241 & 19.0 & 0.691 & \bf{2.97} & 40.6 & 60.4 & - \\
        GemNet-OC-L-F+E \cite{gasteiger2022graph}
            & - & - & - & - & - & - & - & - & 348\\
        \midrule
        {\model{} $L$=6 $K$=16 4-tap 2-band}  & 168M & 414d & \bf{228} & 17.8 & \bf{0.696} & \bf{2.95} & \bf{43.3} & 64.9 & 328 \\
        {\model{} $L$=8 $K$=20}  & 271M & 1280d & 237 & \bf{17.2} & \bf{0.698} & 2.89 & \bf{43.6} & \bf{67.5} & \bf{321} \\
      \bottomrule
    \end{tabular}
    }
    \caption{Comparison of \model{} to existing GNN models on the S2EF, IS2RS and IS2RE tasks when trained on the All or All+MD datasets. Average results across all four test splits are reported. We mark as bold the best performance and close ones, \ie, within 0.5 meV/\AA~MAE, which we found to empirically provide a meaningful performance difference. Training time is in GPU days. Median represents the trivial baseline of always predicting the median training energy or force across all the validation atoms. The SCN $L = 8$ model has $K=20$ layers, $C = 128$ and $E = 256$, while the SCN $L = 6$ model has $K=16$ layers, $C = 128$, $E = 128$ and an energy loss coefficient of 4. }
    \vspace{-0.35cm}
    \label{tab:comp-all}
\end{table*}

We present results on the Open Catalyst 2020 (OC20) dataset \cite{OC20} that is released under a Creative Commons Attribution 4.0 License. OC20 contains over 130M training examples for the task of predicting atomic energies and forces for catalysts used in renewable energy storage and other important applications \cite{zitnick2020introduction}. This dataset is a popular benchmark for the ML community. We begin by comparing results across all tasks on the test set. Next, we show numerous ablation studies comparing model variations on the smaller OC20 2M dataset. Finally, since several papers only report results on the IS2RE task using approaches that directly predict relaxed energies, we also train a model specifically for this task and approach, and compare in Appendix \ref{sec:appx-direct}.

\subsection{Implementation Details}

The implementation of spherical harmonics and their transformations uses the code provided by Euclidean neural networks (e3nn) \cite{e3nn}. For message passing, two resolutions of spherical harmonics are used based on a atom's nearest neighbors, see Appendix \ref{sec:details-res} for exact parameters. Unless otherwise stated $C = 128$, $K=16$, $H = 1024$, $E=128$, and for $\bm{F}_e$ in message passing only orders $m \in [-1, 1]$ are used. All forces are estimated directly as an output of the network, unless stated that the energy conserving gradient-based approach was used. During training, the coefficients for the force and energy losses are 100 and 2 respectively. Training is performed using the AdamW optimizer \cite{loshchilov2017decoupled} with a learning rate of 0.0004. The effective batch size is increased using data parallelism and PyTorch's Automatic Mixed Precision (AMP). All model code will be open sourced with an MIT license in the Open Catalyst Github repo. Please see the supplementary for details on the PaiNN\cite{schutt2021equivariant} baseline.  

\subsection{OC20 Tasks}

We compare against numerous models trained on the OC20 2M, All and MD datasets across the Structure to Energy and Forces (S2EF), Initial Structure to Relaxed Structure (IS2RS) and Initial Structure to Relaxed Energy (IS2RE) tasks \cite{OC20}. OC20 2M, OC20 All and OC20 MD have 2, 133, and 38 million training examples respectively. The 12 and 16 layer SCN models significantly outperform state-of-the-art GemNet-OC \cite{gasteiger2022graph} on S2EF force prediction ($\approx14\%$ improvement) and IS2RE energy predictions ($\approx10\%$ improvement) when trained on OC20 2M, Table \ref{tab:comp-ablation}. Comparable results are found for energy MAE on S2EF with GemNet-OC. We also report throughput efficiency. Note that while the SCN models process fewer samples per second, they are more sample efficient than GemNet (Figure \ref{fig:msg}(b)). Efficiency improvements have also been noted for fully equivariant models \cite{batzner20223}. 

We compare our largest SCN models trained on the All and the All+MD datasets (133M + 38M examples) across all OC20 tasks in Table \ref{tab:comp-all}. For these experiments we use a deep 1-tap model, and a slightly shallower 4-tap model since the use of 4-taps requires more memory. Our model achieves state-of-the-art results for force MAE ($\approx9\%$ improvement) and force cosine for the S2EF tasks. The SCN $L=6$ model performs similarly to the recently released GemNet-OC model \cite{gasteiger2022graph} on energy MAE for S2EF. On the challenging relaxation based IS2RE task that requires models to predict both accurate forces and energies, SCN outperforms GemNet-OC by over $7\%$. Note that we only train a single model to predict both energy and forces, where GemNet-OC-L-F+E + MD is the combination of two models, one trained specifically for forces and the other for energy.

\subsection{Ablation studies}

We explore numerous model variations in Table \ref{tab:comp-ablation} trained on the OC20 2M dataset and evaluated on the OC20 validation set. The first set of ablation experiments explores reducing the complexity of our model. If messages are not rotated before and after applying $\bm{F}_m$ in Equation \ref{eqn:msg}, \ie, $\mathbf{A}_{st}$ is an identity matrix, the results are significantly worse for energies and even more so for forces since forces are highly dependent on angular information. If we replace the non-linear aggregation (Equation \ref{eqn:aggr}) with a simple summation (No 1x1 conv) in a model with $m \in [-1, 1]$, the results also degrade across all metrics. Results improve for $m \in [-1, 1]$ as compared to only using $m=0$ coefficients, which demonstrates that if the equivariant to rotation constraint is relaxed, improved results may be achieved. For higher $m$, $m \in [-2,2]$ or $m \in [-4,4]$ diminishing returns are noticed with increased computational cost.

In some applications, such as molecular dynamics, energy conserving models are needed. We trained an energy conserving model that estimates forces based on the gradients of the energy with respect to the atom positions. Since this approach is more expensive both in memory and computation, we used a smaller model with $L = 4$, $H = 512$ and 12 layers (grad-forces). While the accuracy of the model is good, the results are slightly worse and is 10 times slower than a comparable model that directly estimates forces (direct-forces).

When scaling the network, increasing $L$ has a more significant impact on force estimation, while depth improves energy estimation. This may be due to higher $L$ resulting in greater angular resolution, which is important to force estimation. Greater depth, which allows information to travel further, may lead to better energy predictions since energy is function of the entire atomic structure.   

Sampling multiple z-axis rotations when computing $\bm{F}_e$ with 4-taps (Equation \ref{eqn:taps}) produces significantly improved force predictions, while using two bands ($L$ and $L-1$) when aggregating messages improves both energy and force prediction. The use of 2 bands reduces model throughput a small amount, while sampling with 4-taps reduces throughput by over $2\times$. However, sample efficiency is further increased with 4-taps, resulting in similar training times for the same accuracy, Figure \ref{fig:msg}(b).

\section{Related work}
\label{sec:related}
Molecular and atomic property prediction has made significant recent progress. Models relying on hand-crafted representations like Behler-Parrinello\cite{behler2016perspective} and sGDML\cite{chmiela2018towards} have recently been surpassed by learned feature representations using GNNs \cite{batzner20223, klicpera_dimenetpp_2020, klicpera2021gemnet, schutt2021equivariant}. Early GNN developments focused only on invariant representations. CGCNN\cite{xie2018crystal} and SchNet\cite{schutt2017schnet} make energy and force estimations using only atomic species and pair-wise distance information. DimeNet\cite{klicpera_dimenetpp_2020, klicpera2020directional}, SphereNet\cite{liu2021spherical}, and GemNet\cite{klicpera2021gemnet, gasteiger2022graph} extend this to explicitly capture triplet and quadruplet angles, which are scalars. Utilizing invariant representations offers flexibility in model architecture design since the node or edge features are inherently invariant and model constraints are not needed to maintain equivariance. The enumeration of triplets and quadruplets of atoms to calculate functions of relative or dihedral angles requires careful model design to maintain efficiency. More recently, equivariant models, capturing both scalar and equivariant features, have outperformed traditional GNNs on small molecular datasets including MD17\cite{chmiela2017machine} and QM9\cite{ramakrishnan2014quantum}. While not strictly equivariant, our work falls into this later class of GNNs for atomic property predictions.

Models that strictly enforce equivariance to the SO(3) (3D rotations) or E(3) group (3D rigid transformations) share similarities to our use of spherical channels \cite{schutt2021equivariant,anderson2019cormorant,batzner20223,cohen2018spherical,brandstetter2021geometric}. Both represent atoms by higher-order tensors and not just scalar information. For instance, Tensor Field Networks \cite{thomas2018tensor}, Cormorant \cite{anderson2019cormorant}, NequIP \cite{batzner20223}, SEGNN \cite{brandstetter2021geometric} and BOTNet \cite{batatia2022design} use the same spherical harmonic representations as SCNs, while PaiNN \cite{schutt2021equivariant} is an  $l=1$ variation. All models place constraints on the operations that can be performed to ensure equivarance, such as restricting non-linear operators to invariant scalar inputs, e.g., pair-wise atom distances. In networks like TFN, geometric information of higher degree $l$ are mixed into the $l=0$ features via tensor products on which a gated non-linear function is applied. Similar to our approach, the recently proposed SEGNN uses the spherical harmonic representation to enable steerable features based on edge orientations. SEGNN introduces steerable MLPs that enforce equivariance by restricting the learnable parameters to those that scale the Clebsch-Gordan tensor products followed by gated non-linearities \cite{weiler20183d}. In our approach, we increase the expressivity of the model by identifying a set of $m=0$ coefficients across degrees $l$ that are invariant to rotations given an edge's orientation to which unconstrained functions may be applied. We also demonstrate that improved performance can be achieved by relaxing the requirement for strict message equivariance by including a broader set of coefficients beyond $m=0$. Furthermore, we project the spherical channels onto a grid and perform a pointwise $1 \times 1$ convolution followed by a non-linearity. Although not strictly equivariant, the pointwise operation allows for complex mixing of different degrees of spherical harmonics resulting in rich geometric descriptions. Perhaps as a result of the network's increased expessivity, we see an increase in accuracy with higher degrees ($L=4,6,8$) unlike previous approaches that typically use $L=1$ or $L=2$.

\section{Discussion}
\label{sec:discussion}
A limitation of SCN is that it is very computationally expensive if energy conservation is enforced through the estimation of forces using energy gradients with respect to atom positions. This may limit its use in chemistry applications such as molecular dynamics. Non-energy conserving models are still practically useful for applications such as structure relaxations and transition state searches. The SCN model scales $O(n^2)$ with respect to $L$, which may limit using $L$ larger than 8. Using a larger set of $m$ coefficients during message passing also faces challenges, since the network may find it more difficult to learn approximately equivariant mappings. In Table \ref{tab:comp-ablation}, we saw that increased depth improves results, especially to energy prediction. It remains an open question whether deeper networks will see further improvement.

Given SCN's increased expressivity due to the relaxation of the equivariance constraint and larger model sizes, it is prone to overfitting on smaller datasets. SCN's results on datasets such as MD17\cite{chmiela2017machine} and QM9\cite{ramakrishnan2014quantum} are not state-of-the-art. However, in future work it would be interesting to explore pre-training on larger datasets, such as OC20, and see if improved results can be achieved on smaller datasets through finetuning \cite{kolluru2022transfer}. 

While this work is motivated by problems we face in addressing climate change, advances in chemistry may have numerous use cases. Some of these use cases may have unintended consequences, such as the development of the Haber-Bosch process for ammonia fertilizer production that enabled the world to feed a growing population, but over fertilization has recently led to ocean “dead zones” and its production is very carbon intensive. More alarmingly, the knowledge gained from fertilizer production was used to create explosives during wartime \cite{hager2009alchemy}. We hope to steer research in this area towards beneficial applications by utilizing datasets such as OC20.

In conclusion, we propose a GNN that uses spherical harmonics to represent channel embeddings that contain explicit orientation information. While the network is encouraged to learn output mappings that are equivariant to rotations, best performance is achieved by relaxing the equivariant constraints and allowing for more expressive non-linear transformations. We demonstrate state-of-the-art results across numerous tasks on the large-scale OC20 dataset for modeling catalyst for applications addressing climate change.

\bibliography{reference}

\section*{Checklist}

\begin{enumerate}

\item For all authors...
\begin{enumerate}
  \item Do the main claims made in the abstract and introduction accurately reflect the paper's contributions and scope?
    \answerYes{}
  \item Did you describe the limitations of your work?
    \answerYes{}, see section \ref{sec:discussion}.
  \item Did you discuss any potential negative societal impacts of your work?
    \answerYes{}, see section \ref{sec:discussion}.
  \item Have you read the ethics review guidelines and ensured that your paper conforms to them?
    \answerYes{}
\end{enumerate}

\item If you are including theoretical results...
\begin{enumerate}
  \item Did you state the full set of assumptions of all theoretical results?
    \answerNA{}
        \item Did you include complete proofs of all theoretical results?
    \answerNA{}
\end{enumerate}

\item If you ran experiments...
\begin{enumerate}
  \item Did you include the code, data, and instructions needed to reproduce the main experimental results (either in the supplemental material or as a URL)?
    \answerYes{}, code is released with MIT license, \href{https://github.com/Open-Catalyst-Project/ocp}{OCP GitHub} .
  \item Did you specify all the training details (e.g., data splits, hyperparameters, how they were chosen)?
    \answerYes{}
        \item Did you report error bars (e.g., with respect to the random seed after running experiments multiple times)?
    \answerNo{}, experiments are too expensive to run multiple times. However, we provide rough estimates for run variations in Table \ref{tab:comp-ablation}.
        \item Did you include the total amount of compute and the type of resources used (e.g., type of GPUs, internal cluster, or cloud provider)?
    \answerYes{}, training times and throughput are proved.
\end{enumerate}

\item If you are using existing assets (e.g., code, data, models) or curating/releasing new assets...
\begin{enumerate}
  \item If your work uses existing assets, did you cite the creators?
    \answerYes{}
  \item Did you mention the license of the assets?
    \answerYes{}
  \item Did you include any new assets either in the supplemental material or as a URL?
    \answerNo{}
  \item Did you discuss whether and how consent was obtained from people whose data you're using/curating?
    \answerNA{}
  \item Did you discuss whether the data you are using/curating contains personally identifiable information or offensive content?
    \answerNA{}
\end{enumerate}

\end{enumerate}


\clearpage
\appendix
\section*{Appendix: Table of contents}
\begin{itemize}
    \item[] \ref{sec:appx-equiv} Empirical measurement of SCN's equivariance to rotations
    \item[] \ref{sec:appx-direct} OC20 IS2RE Direct Results    
    \item[] \ref{sec:appx-details} Implementation details
    \item[] \ref{sec:appx-sphharm} Note on spherical harmonics properties
    \item[] \ref{sec:overfit} Overfitting on the training dataset
    \item[] \ref{sec:appx-size} Impact of model size
    
\end{itemize}


\section{Empirical measurement of equivariance to rotations}
\label{sec:appx-equiv}
The SCN is not strictly equivariant to rotations, but depending on the design choices approximate equivarance may be achieved. We begin by empirically measuring the network's invariance and equivariance to rotation for energy and forces respectively. We accomplish this by computing the absolute difference between pairs of results computed from an input and a randomly rotated version of the input:
\begin{equation}
\left| SCN_{energy}(d, a) - SCN_{energy}(\mathbf{R}d, a)\right|,
\label{eqn:energy_equiv}
\end{equation}

\begin{equation}
| SCN_{force}(d, a) - \mathbf{R}^{-1}SCN_{force}(\mathbf{R}d, a)|,
\label{eqn:force_equiv}
\end{equation}
where $d$ is the 3D positional difference between pairs of atoms, $a$ are their atomic numbers, and $\mathbf{R}$ is a random 3D rotation matrix. $SCN_{energy}$ and $SCN_{force}$ are the network's predictions for energy and forces respectively. 

Mean Absolute Difference (MAD) results for various model choices are shown in Table \ref{tab:all-equiv} for models with 12 layers and $L = 6$. Differences are averaged over a model's outputs for a random 1,000 atomic structures. There are four sources that may lead the network to predict different values for rotated versions of the input: 1) the use of $m \neq 0$ coefficients during message passing, 2) non-linear message aggregation, Equation (3), 3) the energy and force output blocks, Equations (4,5), and 4) limits to numerical precision especially when using Automatic Mixed Precision (AMP). 

In Table \ref{tab:all-equiv}, we observe that the MAD increases when $m \in [-1,1]$ coefficients are used during message passing as compared to the $m=0$ message passing that is equivariant. However, the MAD of the $m \in [-1,1]$ model that uses 4 taps during message passing is nearly identical to $m=0$ when not using AMP. This shows that the 4-tap model is close to producing equivariant results. Note MAD is non-zero even for $m=0$ due to the other sources of errors. The $1\times 1$ convolution results in greater rotational differences for forces, while $m \in [-1,1]$ during message passing has a bigger impact on energies. If AMP is used, higher MADs are found due to the limits of numerical precision. 

\begin{table}[h]
    \centering
    \renewcommand{\arraystretch}{1.0}
    \setlength{\tabcolsep}{5pt}
    \resizebox{0.8\linewidth}{!}{
    \begin{tabular}{ll|c|cc|cc}
      \toprule
          & & \mc{5}{|c}{Energy} \\
         \midrule
         & & & \mc{2}{|c}{No AMP} & \mc{2}{|c}{AMP} \\
          \textbf{Model} & & MAE [meV] $\downarrow$ & MAD [meV] $\downarrow$ & $\%$ Error $\downarrow$ & MAD [meV] $\downarrow$ & $\%$ Error $\downarrow$ \\
        \midrule
        {\model{}}  & $m=0$ & 307 & 7.9 & $2.6\%$ & 12.4 & $4.0\%$  \\
        {\model{}}  & $m \in [-1,1]$ & 302 & 39.1 & $12.9\%$ & 42.8 & $14.2\%$ \\
        {\model{}}  & $m \in [-1,1]$, No 1x1 conv & 313 & 30.3 & $9.7\%$ & 37.3 & $11.9\%$ \\
        {\model{}}  & $m \in [-1,1]$, 4-tap & 294 & 6.9 & $2.3\%$ & 19.4 & $6.6\%$ \\
         \midrule
         & & \mc{5}{|c}{Force} \\
         \midrule
         & & & \mc{2}{|c}{No AMP} & \mc{2}{|c}{AMP} \\
         \textbf{Model} & & MAE [meV/\AA] $\downarrow$ & MAD [meV/\AA] $\downarrow$ & $\%$ Error $\downarrow$ & MAD [meV/\AA] $\downarrow$ & $\%$ Error $\downarrow$\\
        \midrule
        {\model{}}  & $m=0$ & 26.5 & 0.4 & $1.5\%$ & 0.9 & $3.6\%$ \\
        {\model{}}  & $m \in [-1,1]$ & 24.6 &  3.1 & $12.6\%$ & 4.4 & $17.9\%$ \\
        {\model{}}  & $m \in [-1,1]$, No 1x1 conv & 26.2 & 0.5 & $2.0\%$ & 0.6 & $2.2\%$ \\
        {\model{}}  & $m \in [-1,1]$, 4-tap & 23.1 & 0.4  & $1.7\%$ & 2.6 & $11.3\%$ \\
      \bottomrule
    \end{tabular}}
    \caption{Mean Absolute Difference (MAD) between pairs of results computed from an input and a randomly rotated version of the same input for energy and forces. The inverse rotation is performed on the second set of forces before comparing results. Mean Absolute Errors (MAE) with respect to ground truth results are shown for reference. Results are shown when using AMP and not using AMP (single precision default). The MAD values as a percentage of the total error is also provided. All values are averaged over 1,000 atomic structures using a model with 12 layers and $L=6$. If a model is equivariant to rotations, the MAD values will be zero.}
    \label{tab:all-equiv}
\end{table}

\subsection{Message aggregation}

Next, we discuss the degree to which our message aggregation Equation (3) is empirically equivariant. If Equation (3) was equivariant, the following would hold:
\begin{equation}
 \bm{G}^{-1}\left(\bm{F}_c\left(\bm{G}(\mathbf{D}\bm{m}_{t}^{(k+1)}), \bm{G}(\mathbf{D}\bm{x}_t^{(k)})\right)\right) = \mathbf{D} \bm{G}^{-1}\left(\bm{F}_c\left(\bm{G}(\bm{m}_{t}^{(k+1)}), \bm{G}(\bm{x}_t^{(k)})\right)\right)
 \label{eqn:aggr_equiv}
\end{equation}
where $\mathbf{D}$ is a block diagonal matrix containing Wigner-D matrices that perform a rotation of the coefficients. That is, the same result should be achieved whether the input or output is rotated. Since the transformation $G$ uses a discrete sampling of the sphere, Equation (3) is clearly not strictly equivariant. The neural network $F_c$ may introduce non-linearities that result in frequencies higher than those able to be represented by the maximum degree $L$ used by the spherical harmonics. As a result, the sampling used by $G$ may be below the Nyquist rate needed to avoid aliasing. This may be minimized but not eliminated by utilizing smooth activation functions such as SiLU. However, in practice we find that the use of Equation (3) with sampling resolution of $2(L+1)$ or higher does in practice lead to results that are approximately equivariant. In Table \ref{tab:aggr-equiv}, we show several empirical results with different depths and activation functions. If only a linear layer is used, the function is equivariant and there is no difference in computed values. For non-linear activations, such as ReLU and SiLU, differences of $4\% - 7\%$ are seen between the rotated versions. Percentages are the Mean Absolute Difference (MAD) divided by the mean absolute value of the outputs. MAD is computed using:
\begin{equation}
 \left|\bm{G}^{-1}\left(\bm{F}_c\left(\bm{G}(\bm{m}_{t}^{(k+1)}), \bm{G}(\bm{x}_t^{(k)})\right)\right) - \mathbf{D}^{-1} \bm{G}^{-1}\left(\bm{F}_c\left(\bm{G}(\mathbf{D}\bm{m}_{t}^{(k+1)}), \bm{G}(\mathbf{D}\bm{x}_t^{(k)})\right)\right)\right|
 \label{eqn:aggr_equiv}
\end{equation}
Minimal differences are seen between using SiLU or ReLU. However, higher percentage differences are observed when using two layers of non-linearities when compared to one. We use two-layers followed by a linear layer in the models for this paper.

\begin{table}[t]
    \centering
    \renewcommand{\arraystretch}{1.0}
    \setlength{\tabcolsep}{5pt}
    \resizebox{0.35\linewidth}{!}{
    \begin{tabular}{ll|c}
      \toprule
         \textbf{Model} & Activation & $\%$ MAD \\
         \midrule
      \midrule
        {\model{}} 1-Layer  & Linear & $0.0\%$  \\
        {\model{}} 1-Layer  & SiLU & $4.6\%$ \\
        {\model{}} 2-Layer  & SiLU & $6.7\%$ \\
        {\model{}} 1-Layer  & ReLU & $4.2\%$  \\
        {\model{}} 2-Layer  & ReLU & $6.8\%$  \\
      \bottomrule
    \end{tabular}}
    \caption{The percentage difference between outputs of message aggregation when rotated. Differences are computed between pairs of examples. The first example is not rotated, the second example's inputs are rotated and the outputs rotated by its inverse, Equation (\ref{eqn:aggr_equiv}). The percentage of the Mean Absolute Differences with respect to the mean absolute values of the outputs are shown. Results are shown for different activation functions and depths. }
    \label{tab:aggr-equiv}
\end{table}

\subsection{Energy and Force Output Blocks}
\label{sec:equiv_efblocks}

Our final set of experiments measure the empirical invariance and equivariance to rotations for the output energy and force blocks. The differences between the not rotated and rotated versions is computed using:

\begin{equation}
\left|\sum_i \int_{\hat{\bm{r}}} \bm{F}_{energy}\left(s_{i}^{(K)}(\hat{\bm{r}})\right) - \sum_i \int_{\hat{\bm{r}}} \bm{F}_{energy}\left(\mathbf{D}s_{i}^{(K)}(\hat{\bm{r}})\right)\right|,
\label{eqn:energy_out_equiv}
\end{equation}

\begin{equation}
\left|\sum_i \int_{\hat{\bm{r}}} \hat{\bm{r}}\bm{F}_{force}\left(s_{i}^{(K)}(\hat{\bm{r}})\right) - \mathbf{R}^{-1}\sum_i \int_{\hat{\bm{r}}} \hat{\bm{r}}\bm{F}_{force}\left(\mathbf{D}s_{i}^{(K)}(\hat{\bm{r}})\right)\right|,
\label{eqn:force_out_equiv}
\end{equation}
where $\mathbf{R}$ is a random 3D rotation matrix and $\mathbf{D}$ its corresponding block diagonal matrix containing the Wigner-D matrices. Results are shown in Table \ref{tab:out-equiv}. For both energy and forces the results are nearly equivariant, and demonstrate the outputs blocks have negligible negative impact on the overall network's equivariance. This is likely due to the force and energy output blocks integrating over the entire sphere, which may result in any differences negating each other. 

\begin{table}[t]
    \centering
    \renewcommand{\arraystretch}{1.0}
    \setlength{\tabcolsep}{5pt}
    \resizebox{0.85\linewidth}{!}{
    \begin{tabular}{ll|cc|cc}
      \toprule
          &  & \mc{2}{c}{Energy}  & \mc{2}{|c}{Force}  \\
         \textbf{Model} & Activation & MAE [meV] & MAD [meV] & MAE Force [meV/\AA] & MAD [meV/\AA]  \\
      \midrule
        {\model{}} 3-Layer  & SiLU & 302 & 0.95 & 24.6 & 0.05 \\
      \bottomrule
    \end{tabular}}
    \caption{Mean Absolute Differences (MAD) for 3-layer energy and force output blocks when the inputs are rotated, Equations (\ref{eqn:energy_out_equiv},\ref{eqn:force_out_equiv}). The MADs are  small ($<1\%$) when compared to the energy and force MAEs, and significantly smaller than other blocks.}
    \label{tab:out-equiv}
\end{table}

\section{OC20 IS2RE Direct Results}
\label{sec:appx-direct}

The task of Initial Structure to Relaxed Energy (IS2RE) may be accomplished using two approaches: 1) Use an S2EF model to relax the positions of the atoms (find local energy minima) and output the energy at the relaxed structure, or 2) directly predict the relaxed energy from the initial structure without performing the relaxation. Empirically, the first approach reported previously in the paper achieves higher accuracy (see Tables \ref{tab:comp-ablation} and \ref{tab:comp-all} in the main paper), while the second approach is more efficient both during training and inference. Although solving the same problem, the second approach of direct prediction is a fundamentally different problem than the first. The direct approach cannot take advantage of detailed position and angular information, since the initial structure only contains a rough placement of the atoms. Instead it must rely more on global knowledge of how the atoms are arranged. Thus, an approach that works well for relaxation based approaches to IS2RE might not work well for direct approaches. 

As shown in Table \ref{tab:IS2RE}, the SCN model achieves state-of-the-art results for energy MAE across all test splits for models that do not use ancillary loses \cite{ying2021transformers,godwin2021simple}. Energy within Threshold (EwT) is comparable to SEGNN \cite{brandstetter2021geometric}. Improvements for SCN are most pronounced for the out-of-domain splits, demonstrating its increased ability to generalize to unseen data. 

\begin{table*}[t]
    \centering
    \renewcommand{\arraystretch}{1.0}
    \setlength{\tabcolsep}{5pt}
    \resizebox{0.91\linewidth}{!}{
    \begin{tabular}{l|ccccc|cccc}
      \mc{1}{c}{} & \mc{9}{c}{OC20 IS2RE Direct Test} \\
      \toprule
        & \mc{5}{c|}{Energy MAE [meV] $\downarrow$} & \mc{4}{c}{EwT [$\%$] $\uparrow$} \\
         Model & ID & OOD Ads & OOD Cat &  OOD Both & Average & ID & OOD Ads & OOD Cat &  OOD Both \\
     \midrule
        Median baseline & 1750 & 1879 & 1709 & 1664 & 0.71 & 0.72 & 0.89 & 0.74 \\
     \midrule
        CGCNN\cite{xie2018crystal} & 615 & 916 & 622 & 851 & 751 & 3.40 & 1.93 & 3.10 & 2.00 \\ 
        SchNet\cite{schutt2018schnet} & 639 & 734 & 662 & 704 & 685 & 2.96 & 2.33 & 2.94 & 2.21 \\
        PaiNN~\cite{schutt2021equivariant} & 575 & 783 & 604 & 743 & 676 & 3.46 & 1.97 & 3.46 & 2.28 \\
        TFN (SE$_{lin}$)~\cite{brandstetter2021geometric} & 584 & 766 & 636 & 700 & 672 & 4.32 & 2.51 & 4.55 & 2.66 \\ 
        GemNet-dT\cite{klicpera2021gemnet} & 527 & 758 & 549 & 702 & 634 & 4.59 & 2.09 & 4.47 & 2.28 \\
        DimeNet++\cite{klicpera_dimenetpp_2020} & 562 & 725 & 576 & 661 & 631 & 4.25 & 2.07 & 4.10 & 2.41 \\
        GemNet-OC\cite{gasteiger2022graph} & 560 & 711 & 576 & 671 & 630 & 4.15 & 2.29 & 3.85 & 2.28 \\
        SphereNet\cite{liu2021spherical} & 563 & 703 & 571 & 638 & 619 & 4.47 & 2.29 & 4.09 & 2.41 \\
        SEGNN\cite{brandstetter2021geometric} & 533 & 692 & 537 & 679 & 610 & \bf{5.37} & 2.46 & \bf{4.91} & 2.63 \\
        
        \midrule 
        {\model{}}  & \bf{516} & \bf{643} & \bf{530} & \bf{604} & \bf{573} & 4.92 & \bf{2.71} & 4.42 & \bf{2.76} \\
        \bottomrule
    \end{tabular}}
    \caption{Results on IS2RE OC20 Test for approaches that directly predict the relaxed energies without performing relaxations and do not use auxiliary losses during training. See Tables \ref{tab:comp-ablation} and \ref{tab:comp-all} for results using relaxation based approaches to IS2RE. Results are shown for two metrics; energy Mean Absolute Error (MAE) and Energy within Threshold (EwT). The SCN model uses $L=6$ with 16 layers and is trained for 21 epochs.}
    \label{tab:IS2RE}
\end{table*}

\section{Implementation details}
\label{sec:appx-details}
In this section, we describe additional implementation details not in the main paper.

\subsection{Varying spherical channel resolution}
\label{sec:details-res}
For atoms that are far away from each other, the messages may be computed using lower resolution spherical channels to reduce memory usage while maintaining similar accuracy. For all results in this paper, the 12 closest neighbors to each atom use the settings described in the paper. For neighbors ranked 13 to 40 by distance, $L$ is reduced by 2 with a minimum value of 4, and H is reduced by a factor of 4. Messages are not calculated for neighboring atoms with ranked distances greater than 40 or a cutoff distance greater than 8 \AA.

\subsection{Point sampling on the sphere for output blocks}
\label{sec:sphere-sample}

For both the energy and force output blocks the integrals in Equations 4 and 5 in the main paper are approximated using points samples on the sphere. Finding a set of evenly distributed points on a sphere for an arbitrary number of points remains an open problem. We use points that are approximately evenly distributed using spherical Fibonacci point sets \cite{hannay2004fibonacci}. Since the density of points does vary slightly across the sphere, we weight each point using a Gaussian weighted counting of nearby points on the unit sphere with $\sigma = 0.5$.

\subsection{PaiNN baseline} 

The PaiNN baseline is our reimplementation of Sch{\"u}tt~\etal\cite{schutt2021equivariant} with the difference that forces are predicted directly from vectorial features via a gated equivariant block instead of gradients of the energy output. This breaks energy conservation but is essential for good performance on OC20.

\section{Note on spherical harmonics properties}
\label{sec:appx-sphharm}

In the paper it is stated that $m=0$ spherical harmonics are invariant to rotations about the z-axis. This is easily seen by looking at the equations for the real spherical harmonics. We show the $Y_{lm}$ equations for up to $l=2$ below, parameterized by $\theta$ (polar) and $\phi$ (longitudinal rotation about the z-axis). Notice that all $m=0$ spherical harmonics are only a function of $\theta$, and thus invariant to changes in $\phi$. 

In Equations \ref{eqn:in1} and \ref{eqn:in2} we state that the $m=-1$ and $m=1$ values are sine and cosine functions of $\phi$. If we integrate or fix $\theta$, we see that the values for $m=\{-1,1\}$ do indeed take the form of Equations \ref{eqn:in1} and \ref{eqn:in2}.

\begin{flalign*}
  & & & & Y_{2,-2}(\theta,\phi) &= \sqrt{\frac{15}{16\pi}}\sin(2\phi)\sin^2\theta  \\
 & & Y_{1,-1}(\theta,\phi) &= \sqrt{\frac{3}{4\pi}}\sin\phi\sin\theta & Y_{2,-1}(\theta,\phi) &= \sqrt{\frac{15}{4\pi}}\sin\phi\sin\theta\cos\theta  \\
 Y_{0,0}(\theta,\phi) &= \sqrt{\frac{1}{4\pi}} & Y_{1,0}(\theta,\phi) &= \sqrt{\frac{3}{4\pi}}\cos\theta & Y_{2,0}(\theta,\phi) &= \sqrt{\frac{5}{16\pi}}(3\cos^2\theta - 1) \\
 & & Y_{1,1}(\theta,\phi) &= \sqrt{\frac{3}{4\pi}}\cos\phi\sin\theta & Y_{2,1}(\theta,\phi) &= \sqrt{\frac{15}{4\pi}}\cos\phi\sin\theta\cos\theta  \\
 & & & & Y_{2,2}(\theta,\phi) &= \sqrt{\frac{15}{16\pi}}\cos(2\phi)\sin^2\theta  \\
\end{flalign*}


\section{Overfitting on the training dataset}
\label{sec:overfit}

\begin{figure}[h]
  \centering
  \includegraphics[width=.99\linewidth]{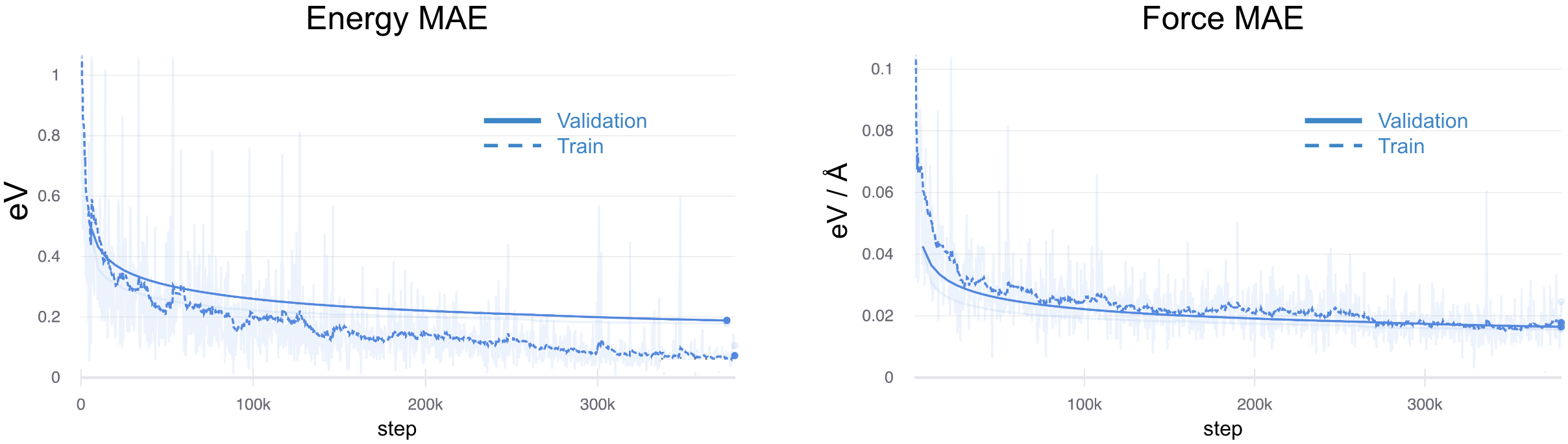}
  \caption{Plot of training and validation errors during training on the \ocd All + MD dataset. The validation error is calculated on a 30k subset of the validation ID dataset. Note significant overfitting is seen for energies but not forces. Errors are smoothed using a 0.9 exponential moving average. Plot is generated from training run in Table \ref{tab:comp-all}}
\label{fig:overfit}
\end{figure}

Figure \ref{fig:overfit} shows the train and validation errors during training. Interestingly, we see very little overfitting on forces, but energy has significant overfitting. This may be due to the energies being a per-structure property, while the forces have more examples since they are a per-atom property. However, overfitting on the energies is still surprising given the large number of examples in the \ocd~All + MD training dataset. This indicates that the SCN model can fit complex functions but could improve on its ability to better generalize through model improvements or data augmentation. Similar trends are found when trained on the \ocd~2M dataset. 

\section{Impact of model size}
\label{sec:appx-size}
\begin{figure}[h]
  \centering
  \includegraphics[width=.99\linewidth]{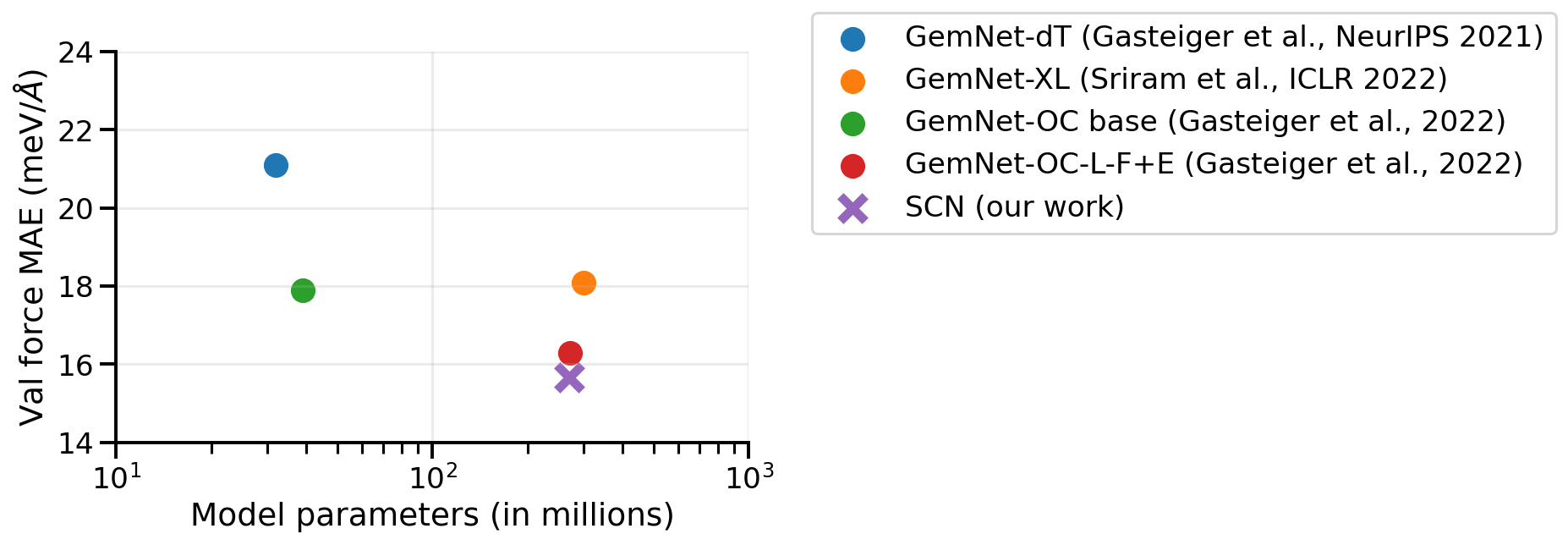}
  \caption{Plot of validation force MAE as a function of model parameters for a variety of models trained on the OC20 S2EF All dataset, including the large variants GemNet-XL and GemNet-OC-L-F+E. Note that for similar model sizes, SCN outperforms the previous state-of-the-art GemNet-OC models.}
\label{fig:params_all}
\end{figure}

\begin{figure}[h]
  \centering
  \includegraphics[width=.50\linewidth]{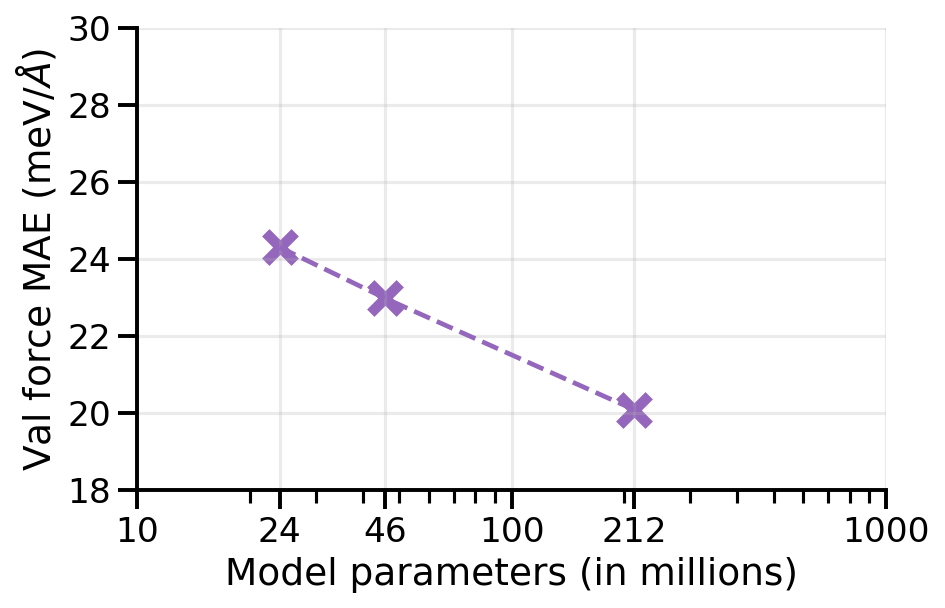}
  \caption{Plot of validation force MAE across three SCN model sizes, all models are trained on the OC20 S2EF 2M dataset.}
\label{fig:params_2M}
\end{figure}

Model size has an impact on the accuracy of the SCN model. In Figure \ref{fig:params_all} we compare the accuracy and model size of SCN and GemNet-OC. For similar model sizes, we see SCN achieves better accuracy than GemNet-OC. This demonstrates that the improvement we see from SCN is more than just the use of larger model sizes. In Figure \ref{fig:params_2M}, we see the accuracy of the SCN model improves as larger models are used on the 2M dataset. However, the accuracy for even the smallest model still outperforms other approaches reported in Table \ref{tab:comp-ablation} in the main paper.

\end{document}